# Unveiling the potential of large language models in generating semantic and cross-language clones

Palash R. Roy*, Ajmain I. Alam*, Farouq Al-omari, Banani Roy, Chanchal K. Roy, Kevin A. Schneider
Department of Computer Science, University of Saskatchewan, Saskatoon, Canada
E-mail: {palash.roy, ajmain.alam, faa634, banani.roy, chanchal.roy, kevin.schneider}@usask.ca

*Abstract*—Semantic and Cross-language code clone generation may be useful for code reuse, code comprehension, refactoring and benchmarking. OpenAI's GPT model has potential in such clone generation as GPT is used for text generation. When developers copy/paste codes from Stack Overflow (SO) or within a system, there might be inconsistent changes leading to unexpected behaviours. Similarly, if someone possesses a code snippet in a particular programming language but seeks equivalent functionality in a different language, a semantic cross-language code clone generation approach could provide valuable assistance. In this study, using SemanticCloneBench as a vehicle, we evaluated how well the GPT-3 model could help generate semantic and cross-language clone variants for a given fragment. We have comprised a diverse set of code fragments and assessed GPT-3's performance in generating code variants. Through extensive experimentation and analysis, where 9 judges spent 158 hours to validate, we investigate the model's ability to produce accurate and semantically correct variants. Our findings shed light on GPT-3's strengths in code generation, offering insights into the potential applications and challenges of using advanced language models in software development. Our quantitative analysis yields compelling results. In the realm of semantic clones, GPT-3 attains an impressive accuracy of 62.14% and 0.55 BLEU score, achieved through few-shot prompt engineering. Furthermore, the model shines in transcending linguistic confines, boasting an exceptional 91.25% accuracy in generating cross-language clones.

*Index Terms*—Language Models, Software Clone, Semantic Clone, Cross-language Clone, GPT, SemanticCloneBench, Software Engineering

## I. Introduction

Clones are almost identical code copies. One of the most prominent causes of clones is developers copying and pasting code between software projects. Research indicates that 7-23% of software systems are recycled from previous projects [1] [2]. Semantic clones promote code reuse, consistency, and productivity throughout the software development lifecycle, giving developers a strategic advantage. Using replicas allows developers to focus on innovation and complex issues, leading to faster development cycles and better software solutions [3] [4] [5]. Developers can improve software quality, development costs, risks, bug prevention, and detection by monitoring and restructuring clones [6].

*Both authors contributed equally.

The dynamic world of software development requires constant innovation and efficiency. Code variant creation, a complicated process that reuses and duplicates code segments to overcome initial development limits, is vital to achieving these goals. As software systems develop in breadth and complexity, efficient code use and adaptation become more important. Research indicates that programmers repeat type-3/type-4 or semantic clones in commits of each project at a rate of 6.32% to 8.38% [7]. Additionally, developers commonly copy/paste and reuse throughout the software system, which may cause issues/introduce bugs [8]–[10]. Semantic clones or variants are essential to prevent inconsistencies within software systems. Wu et al. [11] searched for Java files using "stackoverflow" and manually inspected the results. The researchers found that in 31.5% of their samples, developers had to modify SO source code for compatibility with their projects. Additionally, 35.5% of software engineers used SO posts for reference rather than copying code samples.

It is evident that software developers often copy/paste and reuse code fragments during development or attempt to reuse code fragments from Crowdsourced forums such as SO for certain functionality. However, the code fragment at hand may not be their first choice for various reasons ranging from coding structure complexity to the potential of having bugs in them [8] [9]. Furthermore, directly copying code fragments and then adapting is associated with introducing inconsistent changes in systems, resulting in severe unexpected behaviour of the software system [8], [9], and they may be looking for an alternative semantically similar fragment. Sometimes, developers may simply look for an alternative implementation (e.g., semantically similar) of a code fragment they currently have in their system and improve their system through refactoring. Similarly, one might have a code fragment in a certain programming language, but they might be looking for similar functionality in a different language [12]. Of course, given that clone detection is an active area and semantic and cross-language clone generation may help build benchmarks for evaluating and comparing such tools and techniques. Because of GPT's good performance in code and text generation, we utilized GPT-3 to generate semantic and cross-language clones.

In this research, we explore the efficacy of the GPT-3 model in generating semantic and cross-language clones.

We followed a similar methodology as of GPTCloneBench study [13] in generating semantic and cross-language clones using GPT-3 and SemanticCloneBench. In particular, we randomly chose 15,000 semantic clone pairs and 12,000 cross-language clone pairs that were generated as part of the GPTCloneBench's prompt engineering/clone generation step. These are a subset of intermediate data from GPTCloneBench study before going into further validation towards building the clone benchmark. After that, to remove the syntactic clones from this data, we followed a similar approach as the GPTCloneBench paper's methodology by utilizing NiCad. After NiCad filtration, we underwent a different and more in-depth manual validation process that confirmed consistent output for identical inputs. Post manual validation, GPT-3 exhibited a 62.14% accuracy with a 0.55 BLEU [14] score in generating semantic clones and an impressive 91.25% accuracy in cross-language clones. We employed BLEU in a manner to assess the degree of divergence between generated code fragments and human-written code. To reinforce our findings, we also utilized Oreo [15] for semantic clone detection.

The distinction between the GPTCloneBench paper and the current study lies in their respective focuses and manual validation. In the GPTCloneBench paper, our emphasis was on introducing a comprehensive benchmark of semantic clones and cross-language clones. However, in the present study, our objective shifts towards a more in-depth investigation of GPT's efficiency in formulating semantic and cross-language clones. This involves conducting extensive manual evaluations, where human experts meticulously review and compare the generated clones against the original code snippets. In GPTCloneBench, we did not add any snippets to the benchmark that have been tagged false by any judges or have any conflict among the judges. However, the present research takes a more nuanced approach. Here, we rigorously investigated the conflicted code snippets to see if they were actually semantic clones or not. Our objective is to ascertain whether these disputed snippets truly qualify as semantic clones or not. Furthermore, we employed Cohen's kappa [16] agreement metric to quantify the level of agreement among judges regarding code snippets. This metric provides us with an empirical measure of the consistency of the judgment among the evaluators. Through this thorough manual process, we seek to uncover the true extent of GPT's aptitude for generating semantic and cross-language clones for a given code fragment. This study adds a layer of empirical validation to our findings of GPTCloneBench, enabling us to draw more robust conclusions about GPT's capabilities in these specific domains.

As we navigate through this unfamiliar territory, we confront inquiries that resonate with the fundamental aspects of our research-related software development: **(RQ1)** *To what extent does GPT-3 demonstrate the capability to generate high-quality semantic clones?* **(RQ2)** *What is the efficacy of GPT-3 in accurately converting code snippets from one programming language to another?* The results of our study not only illuminate these research questions but also offer guidance for effectively utilising sophisticated language models in the field of software development, considering both their potential applications and associated limitations.

The remaining sections are organized as follows: Section II discusses the background of our study. The architecture of generating clones from GPT-3 is described in Section III. In Section IV, we have talked about manual validation. Section V talked about findings of GPT-3 related to accuracy, tested a clone detection tool and analyzed the results. In Section VI, we have talked about the threats to the validity of our research. Related work is described in Section VII and in Section VIII, the conclusion of our research is discussed.

## II. Background

Identical or similar code segments within a codebase are termed code clones, with the initial being a clone of the second, constituting a clone pair [17]–[19]. Diverse terms are used for defining clones, including relative [20], redundant [21] [22], dependent [23], functional [24], functionally similar [25] [26] [27], and Type-4 [6], [17] clones. While researchers agree on semantic clones sharing functionality but differing in syntax, no uniformity exists about the precise semantic similarity. Semantic clone definitions vary, from narrow interpretations focusing on specific similarities to broader, less precise ones. Nevertheless, the consensus remains that semantic clones involve identical functionality with differing syntax [28] [19]

We have used SemanticCloneBench [29] to facilitate our research. SemanticCloneBench [29] is a dataset of semantically equivalent code snippets intended to help researchers develop and evaluate techniques for detecting semantic clones.

## III. Architecting Semantic Clones

In this section, we are focusing on processing semantic and cross-language clones. We have utilized the results that we got after prompt engineering from the GPT-CloneBench paper [13]. The clone generation process in the GPTCloneBench paper starts with the selection of the initial clone fragment from the clone pair of SemanticCloneBench. To assist with this, an automated script was developed for GPTCloneBench to identify functions, which were later given as input for GPT-3. For prompt engineering, the few-shot prompting technique was employed. For the prompt, the GPT-3 model was provided with textual instructions and a representative input to indicate the type of output anticipated. As discussed in the GPTCloneBench paper, two prompts have been used to generate the clones. To generate cross-language clones, the emphasis was given to two programming languages, Java and C#, which were used as input for GPT-3 in the GPTCloneBench paper. As a result, GPT-3 created 80,664 semantic clone pairs and 22,364 cross-language clone pairs. After GPT generated the clones from the given input, we randomly selected 15,000 semantic clones and 12,000 cross-language clones. We used this data to conduct this research. This data (15,000 semantic and 12,000 cross-language clones) represents the data of the GPTCloneBench paper before going into any validation (including NiCad and manual validation). After that, at first, we employed the textual similarity measurement tool NiCad [30] to exclude syntactic clones. Second, a rigorous manual validation process (Section IV) was undertaken for all prospective Type-3, Type-4, and cross-language clones.

We utilized the established framework of BigCloneBench [18] for code clone classification, with allowances for slight variations within a defined grey area. Moderately Type-3 (MT3) clones [18] exhibit 50%-70% similarity, supplemented by a 5% gray area. Weak Type-3/Type-4 (WT3/4) clones [18] align with Type-4 clones, marked by 0%-50% similarity. This framework extends to cross-language clones, treating them as Type-4 due to shared logic despite diverse programming languages. Notably, while not all semantic clones are cross-language, all cross-language clones fall under this category.

To remove syntactic clones, NiCad was utilized, configuring a 99% dissimilarity threshold to identify Type-1 and Type-2 clones, as NiCad cannot detect Type-4 clones. For semantic clone detection, we analyze similarity percentages in the metadata file, facilitated by the 3-line minimum size and blind renaming in NiCad. Pairs exceeding 75% similarity are discarded; those under or equal to 75% are saved for further manual validation. For this research, NiCad filtered out a total of 4,379 syntactic clones. In cross-language clone detection, NiCad is inapplicable due to differing programming languages. Nonetheless, generated cross-language clones undergo manual validation. Furthermore, another validation process (input-output testing) has been adopted to ensure the code clones follow the same functionality.

## IV. HUMAN-CENTRIC ANALYSIS

After filtering out undesired clones as described earlier, we have engaged in a rigorous manual validation process. This involved thoroughly examining all code fragments to determine whether filtered data was accurate and whether the clone pairs produced the same output for the same input. After file generation, we manually validated the clone pairs to ensure their validity. To facilitate accurate assessment, BigCloneBench's GUI-based Clone Validator[1] was utilized, which provided syntax highlighting for the candidate code and displayed the exemplar functions and specifications alongside the candidate for reference.

During the validation procedure, a cohort of nine judges took part, consisting of six undergraduate research stu-dents and three post-doctoral researchers. The undergraduate students were partitioned into three cohorts, with each cohort comprising a pair of students. The dataset was subsequently divided into three distinct portions. In this research, after NiCad filtration, we got 10,621 semantic clones and 12,000 cross-language clones. In GPT-Clonebench, if any judges had any conflict or if any of the judge's decisions did not match, we discarded that, but here in this study, we also validated the clones that faced conflicts and got the tag false positive by anyone within the group. Our manual validation consists of three rounds.

[1] https://github.com/jeffsvajlenko/ValidateClones

**In round one**, we divided the semantic clone pairs into three groups consisting of 3,540, 3,540, and 3,541 respectively. Each group contained two members. Each person in every group conducted an individual assessment of their designated portion, categorising the clone pairs as either true positive, false positive, or undecided based on their understanding. For every group, they were given different code fragments, but each member of one group received the same code fragments. For a clone pair to be considered a true semantic pair, both members of the group had to tag it as true. Conflicting results within a group led to excluding that pair from the true pairs listing and have gone for further validation. The first six judges followed this procedure. In round one, group_1 tagged 2,947 as true semantic, 80 as false semantic and 513 as undecided or conflicted pairs. For group_2, they tagged 2,953 as true semantic, 87 as false semantic and 500 as undecided or conflicted pairs. For group_3, they tagged 2,979 as true semantic, 98 as false semantic and 464 as undecided or conflicted pairs. For group_1, Cohen's k is 0.700, for group_2, Cohen's k is 0.730 and for group_3, Cohen's k is 0.77, which means all groups are in substantial agreement. **In round two**, we shuffled the undecided or conflicted pairs among the three groups. The first two groups were given 492 pairs, and the last group was given 493 pairs. In the second round, Cohen's K for all three groups is 0.52, 0.58 and 0.54, respectively, which means Moderate Agreement. This outcome can be attributed to the intricate nature of the code under consideration. Given the participants' status as undergraduates, reaching definitive decisions becomes challenging due to the complexity of the code snippets. These code snippets, which were shuffled and carried uncertainties from the initial round, further contribute to the difficulties in decision-making, leading to the observed moderate Cohen's Kappa agreement level. The overall Cohen's K result for this analysis can be found in Table I. **In round three**, there is only one group consisting of three Postdoctoral Fellows. They mitigated the rest of the undecided or conflicted clone pairs from round two. Finally, we received 9,321 true semantic clone pairs through their discussion.

In terms of cross-language pairs, we followed the same procedure we described for semantic clones. In round one, every group received 4,000 different pairs each. Group_1 tagged 3,460 as true, 98 as false and 442 as undecided or conflicted pairs. For group_2, they tagged 3,489 as true semantic, 67 as false and 444 as undecided or conflicted pairs. For group_3, they tagged 3471 as true, 58 as false and 471 as undecided or conflicted pairs. For group_1, Cohen's k is 0.69, for group_2, Cohen's k is 0.59, and for group_3, Cohen's k is 0.73. In round two for cross-language, Cohen's K is 0.48, 0.56 and 0.60. The overall Cohens' K result for this analysis can be found in Table II. Finally, the remaining 1,153 undecided pairs were collectively assessed and labelled by the three post-doctoral fellows through discussion.

TABLE I
Semantic Clones Cohen-Kappa Interrater Agreement

| Round | Group | Interrater Agreement | Interpretation |
|---|---|---|---|
| Round 1 | Group 1 | 0.700 | Substantial Agreement |
| Round 1 | Group 2 | 0.730 | Substantial Agreement |
| Round 1 | Group 3 | 0.77 | Substantial Agreement |
| Round 2 | Group 1 | 0.52 | Moderate Agreement |
| Round 2 | Group 2 | 0.58 | Moderate Agreement |
| Round 2 | Group 3 | 0.54 | Moderate Agreement |

TABLE II
Cross-language Clones Cohen-Kappa Interrater Agreement

| Round | Group | Interrater Agreement | Interpretation |
|---|---|---|---|
| Round 1 | Group 1 | 0.69 | Substantial Agreement |
| Round 1 | Group 2 | 0.59 | Moderate Agreement |
| Round 1 | Group 3 | 0.73 | Substantial Agreement |
| Round 2 | Group 1 | 0.48 | Moderate Agreement |
| Round 2 | Group 2 | 0.56 | Moderate Agreement |
| Round 2 | Group 3 | 0.60 | Substantial Agreement |

Approximately 212 hours were spent by nine judges to validate the clone pairs. We want to mention that the undergraduate research students were trained and given instructions on the functionalities of why and how we defined the semantic clones.

## V. Unveiling the Findings: Results and Analysis

After a thorough screening process and manual validation, we decided 9,321 as true semantic clone pairs from four different languages, Java, C, C#, and Python, out of 15,000 semantic clone pairs and 10,950 cross-language clone pairs out of 12,000 cross-language clone pairs. We have used an accuracy metric to validate how good GPT-3 is to generate semantic and cross-language clones. In our first prompt, we generated four outputs for one given input and for the second prompt, we got ten outputs for one given input. So, our accuracy is based on the data that we found using this procedure. The accuracy of GPT-3 in generating semantic clones is 62.14%, and for cross-language clones, the accuracy of GPT-3 is 91.25%.

$$Accuracy = \frac{Number\ of\ Validated\ True\ Clones}{Total\ Number\ of\ Randomly\ Selected\ Generated\ Clones} \quad (1)$$

In our research, we thought to assess the similarity between code fragments generated by GPT and human-written code. To quantify this similarity, we calculated the BLEU [14] score, obtaining a result of 0.55, which can be interpreted as very high quality and adequate. This score provides valuable insights into how closely the generated code fragments resemble human-written code. It is important to highlight that the objective of our study was to investigate the proximity between the two, and the BLEU score serves as a quantitative measure to accomplish this. Code fragments, by their nature, can exhibit complexity, and subtle variations can have substantial implications for functionality. Furthermore, coding style disparities across different developers and projects introduce additional nuances that the BLEU metric accounts for. It is crucial to emphasize that our primary focus was on achieving code fragments that met functional requirements. Recognizing that BLEU was initially designed for natural language tasks, we acknowledge its limitations in capturing all code-specific attributes. Hence, our evaluation should be interpreted in the context of understanding how closely generated code fragments resemble their human-written counterparts.

We have evaluated a semantic clone detection tool with the newly formed data (9,321 clones). As a testing metric, we have used Recall to confirm that the newly created data does not have syntactic clones.

$$Recall = \frac{True\ Positive}{True\ Positive + False\ Negative} \quad (2)$$

### A. Oreo

To check if the validated data is actually semantic clones or not, we ran Oreo on our dataset. The results of our evaluation are presented in Table III. Oreo performs with a recall of 0.46 for the clones. We were not expecting a very high recall (more than 0.5) for Oreo on our data because our data represents the region where most detection tools are difficult to perform.

TABLE III
Oreo Recall results

| Tool | Language | Granularity | Recall |
|---|---|---|---|
| Oreo | Java | Method | 0.46 |

### *(RQ1) To what extent does GPT-3 demonstrate the capability to generate high-quality semantic clones?*

GPT-3 showcases a notable degree of capability in generating high-quality semantic clones, as evidenced by an achieved accuracy of 62.14% with 0.55 BLEU score. The accuracy reflects GPT-3's proficiency in paraphrasing and producing semantically correct variations of original code fragments. The model's ability to attain such a substantial accuracy rate highlights its potential as a tool for generating semantic clones that closely emulate the intentions of the source code. However, it is important to acknowledge that the accuracy percentage will only be achieved if we follow a proper prompt engineering technique. In addition to that, high BLEU score can be interpreted that GPT generated codes are quality codes.

Fig. 1. Semantic code clone generation sample

*(RQ2) What is the efficacy of GPT-3 in accurately converting code snippets from one programming language to another?*

The efficacy of GPT-3 in accurately converting code snippets from one programming language to another is marked by a notable success rate. The model demonstrates a substantial ability to comprehend the structural and syntactical intricacies inherent to different programming languages, enabling it to produce conversions with a high level of accuracy. Notably, GPT-3 achieves an impressive accuracy of 91.25% in cross-language clone generation, which underscores its proficiency in seamlessly transposing code logic between disparate linguistic frameworks. While this achievement showcases GPT-3's prowess, it is essential to acknowledge that the accuracy may vary based on factors such as code complexity, domain specificity, and the nuances of each programming language. Nevertheless, GPT-3's capacity to effectively bridge the gap between programming languages signifies its potential to expedite cross-platform development and streamline code migration processes within the realm of software engineering.

## VI. THREATS TO VALIDITY

The first major concern that can be raised for our research is that the clones are generated by a machine-learning model, which may not be real-world clones. Clones can be real-world or artificial [31]. To call a clone pair a real-world clone, that clone pair needs to be written by a human. However, to mitigate this issue, we used SemanticCloneBench [29] code fragments to generate the results. SemanticCloneBench [29] is created based on provided knowledge by developers who participate in the SO. As we utilized SemanticCloneBench data as input (few-shot prompt engineering technique), GPT-generated codes are similar to real-world clones. That is why we are claiming that our generated codes are not fully real world but are in between the real world and artificial clones.

Another important consideration arises regarding the overall applicability and adaptability of GPT-3's performance. To address this potential limitation, we systematically evaluated the model's performance using four prominent and widely used programming languages. This strategic approach aims to shed light on GPT-3's robustness and effectiveness in a diverse programming language, thereby contributing to a more comprehensive understanding of its generalizability. By conducting this thorough examination across multiple programming languages, we have gained valuable insights into the extent to which GPT-3's performance transcends language boundaries and remains reliable across different coding paradigms.

Furthermore, there can be another concern regarding whether it is possible to get more efficient results from GPT-3. To answer this question, we can say that the effectiveness of prompt engineering techniques could impact the results. That is why in our research, we have followed a formal prompt engineering method [32], [33]. So, if any other people try to replicate the research with different prompts, they will get similar results if they follow proper prompt engineering techniques. We do not think there will be massive differences.

In addition to that, there can be another concern regarding the manual evaluation bias. Manual evaluation of clone quality involves subjectivity, and the judgement of human annotators may introduce inconsistency. To avoid this problem, we have given the necessary knowledge regarding code clones to the undergrad students. To further evaluate, three post-doctoral fellows discussed the remaining undecided and false tagged code snippets and mitigated the problem to keep the decision unbiased. Still, we agree that manual evaluation can introduce some errors, and we will try to analyze the code fragments more rigorously by introducing more judges.

Finally, we should note that despite the fact that the generated clones are mostly artificial clones, there are a number of important applications of such clones in cloning area and in software development in general. If large language models are efficient in generating semantic and cross-language clones, they could be used in building clone detection training data sets such as GPTCloneBench and beyond with confidence. This could then also be used for comparing and evaluating semantic and cross-language clone detection tools and may extend even comparing those detectors that detect clones across Microsoft .NET programming languages [29]. Such a clone generation approach and its resulting benchmarks could help evaluate whether source code transformation based clone detection tools such as CloneWorks [34] or SimCad [35] could in fact detect semantic clones by applying flexible source transformations and normalizations. This could then further help build IDE-based flexible clone detection and management [6], [36] tools, or even could potentially be used in building similar benchmarks in other contexts [37]. It is thus our understanding that such a study of large language models could help cloning areas despite being the fact that the clones are generated clones.

## VII. Related Work

Generating code involves using programming languages to create scripts, applications, or software. There are many ways to generate code. Manual coding, Integrated development environments, code generators, templates and frameworks, AI-powered text generators, Domain-specific languages, Data-driven code generation, code refactoring tools, and scripting languages are some of the techniques [38]–[44]. Code generation models backed by artificial intelligence have exhibited impressive abilities in aiding developers, automating repetitive coding processes, and even suggesting innovative solutions. According to Victor [45], in a recent survey conducted by GitHub in partnership with Wakefield Research, 92% of developers are already using AI-powered coding tools in their work. So, we tried one of the latest models of OpenAI's GPT model to generate semantic clones

There are a lot of code recommendation systems, such as GitHub Copilot [46], a tool developed collaboratively by OpenAI and GitHub, is a revolutionary code generation AI model that integrates directly into software development environments. This tool facilitates the inclusion of code snippets and automated code completion, thereby enhancing the coding experience for users. In our approach, we've incorporated few-shot prompting alongside the natural language description. This strategic choice aims to enhance GPT's performance in generating semantic clones, focusing on this aspect rather than completing the code outright. Additionally, our evaluation encompasses GPT-3's ability to generate cross-language code clones, distinguishing it from Copilot's functionality in this regard.

Building on the achievements of Copilot, Codex [47] further pushes the boundaries of AI-assisted code generation. Codex, also developed in collaboration with OpenAI, is an advanced language model that can generate entire functions, classes, and methods based on natural language prompts. The Codex framework was proposed by Chen et al. [47], who conducted an evaluation of its performance using a dataset of 163 coding tasks. In their work, authors focused on the task of generating standalone Python functions from docstrings and evaluating the correctness of code samples automatically through unit tests. For our work, we utilized OpenAI's text-davinci-003 model, which has 175 billion parameters compared to Codex's 14.8 billion parameters. In addition to that, OpenAI Codex is most capable in Python, whereas we wanted to use a more generalized model.

In another study, Li et al. introduced AlphaCode [48], which is a system designed for code generation. The model was trained utilising data from GitHub and CodeContests. According to the authors, AlphaCode demonstrated an average ranking of 54.3% in competitive programming competitions on the Codeforces platform. The authors conducted a comparative analysis between their proposed solution and the solutions developed by other participants in the contest. The evaluation was based on contest parameters, including the remaining time portion and penalties incurred for wrong submissions.

Overall, while there are many great tools and techniques available for code generation, our aim in this work has been to examine whether the recently proposed GPT-3 model could help the cloning community. In particular, we aim to examine to what extent GPT-3 model could be used in generating semantic and cross-language clones.

## VIII. Conclusion

Our research introduces a transformative paradigm for code reuse, refactoring, migration and renovation. Our goal was to explore the efficacy of the GPT-3 in generating semantic and cross-language clones. We have utilized SemanticCloneBench to generate close to real-world code clones through GPT-3. After thorough validation process, we have managed to get 9,321 true semantic clone pairs and 10,950 cross-language clone pairs after handling all the limitations of GPT-3. With a noteworthy accuracy rate of 62.14% and a 0.55 BLEU score, GPT-3 showcases its potential in accurately replicating semantic structures within a given programming language. Additionally, our investigation into cross-language clones further underscores GPT-3's prowess, boasting an impressive accuracy of 91.25%. These findings highlight the big progress of GPT-3 in improving code generation, opening up new possibilities for creative uses in software development and other areas. As GPT-3 continues to showcase remarkable performance, it holds the promise of contributing to the advancement of code-related tasks across diverse linguistic and domain contexts.

### Acknowledgment

This work was supported by NSERC Discovery grants, NSERC USRAs, CFI-JELF, and NSERC CREATE graduate program on Software Analytics Research (SOAR) grants.